\def\niig{\hat{n}_i\cdot\nabla}
\def\njjg{\hat{n}_{i-1}\cdot\nabla}
\def\nkkg{\hat{n}_{i+1}\cdot\nabla}
\def\tiig{\hat{\tau}_i\cdot\nabla}
\def\tjjg{\hat{\tau}_{i-1}\cdot\nabla}
\def\tkkg{\hat{\tau}_{i+1}\cdot\nabla}

\newcommand{\LB}[1]{\label{#1}}

\newcommand{\YY}[1]{\typeout{YY: {#1}}}

\newcommand{\be}{\begin{equation}}
\newcommand{\ee}{\end{equation}}
\newcommand{\bea}{\begin{eqnarray}}
\newcommand{\eea}{\end{eqnarray}}

\documentclass{elsart}
\usepackage{epsfig}
\begin{document}

\begin{frontmatter}



\title{Induced Defect Nucleation and Side-Band Instabilities in Hexagons with
Rotation and Mean flow}


\author{Yuan-Nan Young\thanksref{Co}}
\author{\and Hermann Riecke}

\address{
Department of Engineering Sciences and Applied Mathematics, 
Northwestern University,2145 Sheridan Rd, Evanston, IL, 60208, USA}

\thanks[Co]{Corresponding author. 
email: young@statler.esam.nwu.edu}

\begin{abstract}
The combined effect of mean flow and rotation on hexagonal patterns is
investigated using Ginzburg-Landau equations that include nonlinear 
gradient terms as well as the nonlocal coupling provided by the mean flow.
Long-wave and short-wave side-band instabilities are determined.  Due to
the nonlinear gradient terms and enhanced by the mean flow, the
penta-hepta defects can become unstable to the induced nucleation of
dislocations in the defect-free amplitude, which can lead to the proliferation 
of penta-hepta defects and persistent spatio-temporal chaos. For individual
penta-hepta defects the nonlinear gradient terms enhance climbing or gliding
motion, depending on whether they break the chiral symmetry or not.

\end{abstract}


\end{frontmatter}

\section{Introduction}
\LB{sec:intro}

Patterns in Rayleigh-B\'enard convection have been extensively explored,
both theoretically and experimentally, as paradigms to investigate
spatio-temporal chaos and transitions from ordered to disordered states.
Arguably, one of the most interesting chaotic states  is that of spiral
defect chaos. For small Prandtl numbers, it arises from roll convection
at moderate heating rates in systems with large aspect ratio \cite{MoBo93}.
In this complex chaotic state spirals, disclinations, and dislocations
are  persistently created and annihilated. It  is found to be driven by
a large-scale mean flow,  which arises due to the curvature of the
convection rolls and becomes prominent for small Prandtl numbers.

Another classic chaotic state that arises from a planform of convection rolls is
the domain chaos resulting from the K\"uppers-Lortz instability  in rotating
convection \cite{KuLo69,ZhEc91,HuPe98}. In this instability rolls are
unstable to rolls with a similar wavelength but different orientation. The
resulting state is characterized by domains of convection rolls with different
orientation that persistently invade each other. 

Motivated by the strong impact that mean flow and rotation have on the stability
of rolls 
and by the interesting chaotic states that result from it, we consider
in this paper the effect of mean flow and rotation on the stability and
dynamics of hexagon patterns and their defects. Hexagon patterns are common phenomena in various
pattern-forming systems and are readily obtained also in 
B\'enard-Marangoni convection driven by surface tension (e.g.
\cite{NiTh95,ScVa95}) and in buoyancy-driven non-Boussinesq convection
(e.g. \cite{BoBr91}). In both of these convection systems mean flow 
is expected to be important for small and moderate Prandtl numbers.

In the absence of mean flows, the side-band instabilities and dynamics of
hexagons with broken chiral symmetry (e.g. by rotation) have been
investigated in the context of three coupled Ginzburg-Landau 
equations \cite{EcRi00} as well as in a long-wave model for Marangoni
convection \cite{MaRi02} and a model of Swift-Hohenberg type
\cite{SaRi00}. Even on the level of the lowest-order Ginzburg-Landau description
rotation makes the system non-variational, and oscillatory amplitude
\cite{Sw84,So85,EcRi00a} and side-band instabilities appear
\cite{EcRi00}. If the nonlinear gradient terms are retained in the 
Ginzburg-Landau equations the latter can lead to an interesting state of
spatio-temporal chaos
\cite{EcRi00}.

In the absence of rotation, mean flow couples differently to the
two steady, long-wave phase modes of the weakly nonlinear hexagon patterns
\cite{YoRi02}; only the stability limits due to the transverse
phase instability are affected by the mean flow, while those corresponding to 
the longitudinal
phase mode are unchanged. As a result, for sufficiently small Prandtl
numbers the stability limit for large wavenumbers is given by
the  transverse phase mode, while for small wavenumbers the 
longitudinal mode becomes dominant. This is particularly interesting,
since without mean flow the longitudinal mode is usually of little
importance \cite{SuTs94,YoRi02}. While the
transient patterns arising from the longitudinal instability remain quite
ordered, those  ensuing from the instability involving the transverse mode
typically exhibit domains of hexagons with quite different orientation 
\cite{YoRi02}. 

In this paper, we investigate the combined effects of rotation and of mean
flow on weakly nonlinear hexagonal patterns using appropriately extended
Ginzburg-Landau equations. We retain all three possible cubic non-linear
gradient terms. The nonlinear
evolution of the side-band instabilities leads to the formation of dislocations that later
combine  to penta-hepta defects. In parameter regimes in which the
nonlinear  gradient terms, rotation, and mean flow are significant we find
quite intriguing defect dynamics:  the presence of penta-hepta defects
induces the nucleation of a dislocation pair in the defect-free amplitude,
which can lead to the proliferation of defects and to persistent
spatio-temporal chaos. We further study the effect of the nonlinear gradient
terms on the motion of single penta-hepta defects by calculating their
mobility  and the Peach-K\"ohler-type force acting on them.

This paper is structured as follows. In Section \ref{sec:amplitude} we
formulate the problem by extending previous work on mean flow in hexagons
\cite{YoRi02} to include the breaking of the chiral symmetry by rotation. Then
we investigate in Section \ref{sec:linear01} the linear stability of hexagons
with respect to side-band perturbations.   We demonstrate the induced defect
nucleation and the  resulting spatio-temporal chaos in Section
\ref{sec:num}.   In Section \ref{sec:def} we investigate the effects  of the
nonlinear gradient terms on the motion of isolated penta-hepta defects both 
analytically and numerically.  We discuss our findings in Section
\ref{sec:con}.

\section{Amplitude equations}
\LB{sec:amplitude}

For small Prandtl numbers and with rotation the usual Ginzburg-Landau
equations for the three complex amplitudes $A_j$ making up the hexagon
patterns need to be extended in two ways. The mean flow, which is 
driven by long-wave modulations of the convective
amplitude, provides a nonlocal coupling of the roll modes
  \cite{SiZi81}. Rotation breaks the chiral
symmetry. This is reflected in the difference between the cubic term  
coupling $A_1$ to $A_2$ and that coupling $A_1$ to $A_3$
\cite{Sw84,So85,EcRi00}. Furthermore, to cubic order it introduces an
additional nonlinear gradient term in the  equation for the amplitudes and an
additional term in the leading-order equation for the mean flow $Q$ 
\cite{CoMa00}. The equations for the three hexagon modes in rotating 
convection at finite Prandtl numbers thus read  
\begin{eqnarray}
\LB{eq:mf1}
\partial_t A_j &=&
\left(\mu+ (\hat{n}_j\cdot\nabla)^2 -
|A_j|^2-(\nu-\gamma) |A_{j-1}|^2-(\nu+\gamma)|A_{j+1}|^2\right) A_j \nonumber\\
&& +A_{j-1}^*A_{j+1}^* - i \beta A_j (\hat \tau_j\cdot \nabla) Q \nonumber\\
&& +i(\alpha_1-\alpha_3) A_{j-1}^*(\hat{n}_{j+1}\cdot\nabla)A_{j+1}^*
   +i(\alpha_1+\alpha_3) A_{j+1}^*(\hat{n}_{j-1}\cdot\nabla)A_{j-1}^* \nonumber \\
&& +i\alpha_2(A_{j-1}^*(\hat{\tau}_{j+1}\cdot\nabla)A_{j+1}^*-
              A_{j+1}^*(\hat{\tau}_{j-1}\cdot\nabla)A_{j-1}^*),\\
\LB{eq:mf2}
\nabla^2 Q &=& \sum_{i=1}^3 \left[ 2(\hat{n}_i\cdot\nabla)(\hat{\tau}_i\cdot\nabla)+
               \tau\left((\hat{n}_i\cdot\nabla)^2-(\hat{\tau}_i\cdot\nabla)^2\right)\right]|A_i|^2,
\end{eqnarray}
with $j=1,2,3$ cyclically permuted in (\ref{eq:mf1}) and $\hat{n}_j$ and
$\hat{\tau}_j$ denoting unit vectors parallel and perpendicular to the critical
wave-vector associated with amplitude $A_j$, respectively.  With respect to
the rotation rate, the coefficients  $\gamma$, $\alpha_3$, and $\tau$ are odd
functions, while the other coefficients are even. They also depend on other
physical parameters such as the Prandtl number. Note that in the presence of rotation 
not only  the difference  $((\hat{n}_i\cdot\nabla)^2 -(\hat{\tau}_i\cdot\nabla)^2)|A_i|^2$ 
but also the corresponding sum is allowed by symmetry, which
corresponds to $\nabla^2|A_i|^2$ and leads to a contribution $Q_l \propto
\sum_{i=1}^3|A_i|^2$ to $Q$. Upon insertion in (\ref{eq:mf1}) $Q_l$ contributes to the 
cubic nonlinear gradient terms and provides therefore a local rather
than a non-local coupling of the roll modes. In this paper we focus on the non-local
coupling and neglect this term along with the other cubic gradient terms.

The above equations allow for several stationary, spatially periodic patterns. 
Rolls with amplitudes $A_j=\sqrt{\mu-q^2} e^{i q \hat{n}_j\cdot{\bf x}}$,
$A_{j\pm 1} =0$, exist for $\mu \ge q^2$, and are stable to homogeneous
perturbations for $\mu \ge \mu_R+q^2$, where
$\mu_R=(1+2q\alpha_1)^2/(1-\nu)^2$. Hexagon solutions with
wavenumber slightly different than the critical value $q_c$ are given by 
$A_j = R_0 e^{i q \hat{n}_j\cdot{\bf x}}$ with 
\begin{equation}
\label{eq:amp}
R_0 = \frac{(1+2q\alpha_1)\pm 
\sqrt{(1+2q\alpha_1)^2+4(\mu-q^2)(1+2\nu)}}{2(1+2\nu)}.   
\end{equation}
We will consider only such equilateral hexagons, 
for which the wavenumbers in all
three modes are equal. Non-equilateral  hexagon patterns have been
discussed in the absence of mean flow and rotation in \cite{NuNe00}. Since
the quadratic coupling coefficient in equation (\ref{eq:mf1}) has been scaled
to $+1$, the hexagon solution corresponding to the amplitude $R_0$ with a
minus sign in front of the square root is always unstable. In the following we
will consider only the hexagon solution of amplitude $R_0$ with the plus
sign in equation (\ref{eq:amp}).  It is stable to homogeneous
perturbations if both $u\equiv R_0^2(1-\nu)+(1+2\alpha_1 q)R_0 \ge 0$ and
$v\equiv 2R_0^2(1+2\nu)-(1+2\alpha_1 q)R_0 \ge 0$. Mixed modes with
amplitudes $A_j=1/(\nu-1)$ and $A_{j\pm 1} =\sqrt{(\mu-q^2-A_j^2)/(1+\nu)}$
also exist but are always unstable with respect to rolls or hexagons. The
stability of these stationary, spatially periodic solutions with respect to
homogeneous amplitude perturbations is summarized in the bifurcation
diagram in Fig.\ref{fig:bif}: 
the hexagons first appear at $v=0$ (or $\mu=\mu_{SN}$ in the
bifurcation diagram) through a saddle-node bifurcation, and become
unstable to `oscillating hexagons' via a Hopf bifurcation at $u=0$ (or
$\mu=\mu_{H}$ in Fig.\ref{fig:bif}) with a Hopf frequency
$\omega_c=2\sqrt{3}\gamma(1+2q\alpha_1)^2/(\nu-1)^2$. Here we focus on 
the instability of steady hexagons in the range $\mu_{sn} \le \mu \le \mu_{H}$; a
detailed discussion of the oscillating hexagons for $\mu \ge \mu_{H}$  can
be found in \cite{Sw84,So85,EcRi00,EcRi00b}.

\begin{figure}
\centerline{\epsfxsize=07.0cm\epsfbox{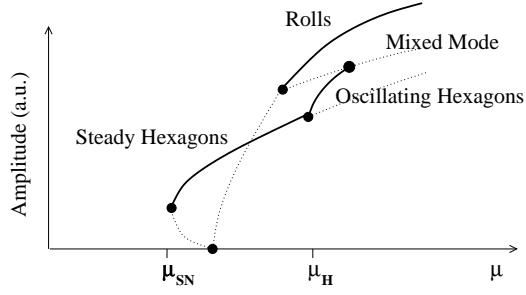}}
\caption{
Sketch of bifurcation diagram of simple, spatially periodic solutions 
of (\ref{eq:mf1},\ref{eq:mf2}) with rotation.
}
\LB{fig:bif}
\end{figure}     

\section{Phase equation and general stability analysis}
\LB{sec:linear01}
Following the procedures in \cite{Ho95,EcRi00}, we first derive the phase
equation that  describes the long-wave side-band instabilities of hexagons.  
The slightly perturbed hexagon solution
\begin{equation}
A_j=R_0 e^{i q \hat{n}_j\cdot{\mathbf x}} (1 + r_j + i \phi_j),\;\;
j=1,\;2,\; 3
\LB{eq:perth}
\end{equation}             
is substituted into  equations (\ref{eq:mf1},\ref{eq:mf2}), where $r_j$ and $\phi_j$ 
are the small amplitude and phase perturbations, respectively. We introduce
the super-slow scales $\partial_t \rightarrow \delta^2\partial_t$ and
$\nabla\rightarrow \delta\nabla$ with $|\delta|\ll 1$, and adiabatically
eliminate the perturbations in  the amplitudes  and in the total phase
$\Phi\equiv \phi_1+\phi_2+\phi_3$ by expressing them in terms of the two
translation phase modes $\phi_x\equiv -(\phi_2+\phi_3)$  and
$\phi_y\equiv(\phi_2-\phi_3)/\sqrt{3}$. At linear order in $\delta$, the
mean-flow amplitude $Q$ can be written in terms  of the phase vector
$\vec{\phi}=(\phi_x,\phi_y)$ as
\begin{eqnarray}
\LB{eq:linear-q}
Q&=&d_{\times 2} \nabla\cdot\vec{\phi} + d_{\perp} \hat{\bf e}_z\cdot\nabla\times\vec{\phi}, 
\end{eqnarray}
where
\begin{eqnarray}
d_{\perp}     &=& \frac{3}{4u^2+\omega^2}[(u-\frac{\omega}{2}\tau) w
                                                +(\tau u +\frac{\omega}{2})\sqrt{3}\alpha_3 R_0^3],\\
d_{\times 2 } &=&\frac{-3}{4u^2+\omega^2}[(\tau u + \frac{\omega}{2}) w
                                                -(u-\frac{\omega}{2}\tau)\sqrt{3}\alpha_3 R_0^3], \\
 w &=& 2qR_0^2+(\alpha_1+\sqrt{3}\alpha_2)R_0^3, \\
\omega &=& 2\sqrt{3}R_0^2\gamma,
\end{eqnarray}                                                     
and $\hat{\bf e}_z$ is the unit vector perpendicular to the $x-y$ plane
(following a right-hand rule). The linearized phase equation for $\vec{\phi}$
thus reads
\begin{eqnarray}
\LB{eq:phase00}
\partial_t\vec{\phi} &=& D_{\perp}^0 \nabla^2\vec{\phi} +
                        (D_{\parallel}^0-D_{\perp}^0)\nabla(\nabla\cdot\vec{\phi}) -
                         D_{\times 1}^0(\hat{\bf e}_z\times \nabla^2 \vec{\phi}) +
                         D_{\times 2}^0(\hat{\bf e}_z\times \nabla)(\nabla\vec{\phi}) \nonumber \\
                     & & -\beta\nabla\times(Q\hat{\bf e}_z),
\end{eqnarray}                                                     
where the coefficients with superscript $0$
correspond to the infinite Prandtl number case ($\beta = 0$) \cite{EcRi00}:
\begin{eqnarray}
D_{\perp}^0 &=& \frac{1}{4} + \frac{1}{4u^2+\omega^2}\left\{ \frac{R_0^2u}{2}\left[(
            \alpha_1+\sqrt{3}\alpha_2)^2+3\alpha_3^2\right]-\sqrt{3}R_0\omega\alpha_3q-2uq^2\right\},\nonumber\\
D_{\parallel}^0 &=& D_{\perp}^0 + \frac{1}{2} - \frac{1}{v}
           \left\{R_0^2\alpha_1(\alpha_1-\sqrt{3}\alpha_2)-R_0(3\alpha_1-\sqrt{3}\alpha_2)q+2q^2\right\},\nonumber\\
D_{\times 1}^0 &=& \frac{1}{4u^2+\omega^2}\left\{\frac{1}{4}\omega R_0^2 
            \left[(\alpha_1+\sqrt{3}\alpha_2)^2+3\alpha_3^2\right] 
            +2\sqrt{3}R_0u\alpha_3q-\omega q^2 \right\},\nonumber\\
D_{\times 2}^0 &=& \frac{\alpha_3}{v}\left(\sqrt{3}R_0^2\alpha_1-\sqrt{3}R_0q\right).
\LB{eq:diff0}
\end{eqnarray}                                                       
Substituting equation (\ref{eq:linear-q}) into the linear phase equation (\ref{eq:phase00}),
we obtain 
\begin{equation}
\LB{eq:phase}
\partial_t\vec{\phi} = D_{\perp} \nabla^2\vec{\phi} + 
                      (D_{\parallel}-D_{\perp})\nabla(\nabla\cdot\vec{\phi}) - 
                      D_{\times 1}(\hat{\bf e}_z\times \nabla^2 \vec{\phi}) +
                      D_{\times 2}(\hat{\bf e}_z\times \nabla)(\nabla\vec{\phi}),
\end{equation}
where $D_{\perp} = D_{\perp}^0+\beta d_{\perp}$, 
$D_{\parallel}=D_{\parallel}^0$,
$D_{\times 1} = D_{\times 1}^0$ and 
$D_{\times 2} = D_{\times 2}^0+\beta d_{\times 2}$. The terms
$\beta d_{\perp}$ and $\beta d_{\times 2}$ are the mean flow
contributions to the diffusion coefficients.

The two eigenvalues for normal mode solutions to equation (\ref{eq:phase}) are 
\begin{equation}
\sigma_{\pm} = -\frac{\sf k^2}{2}\left[ D_{\parallel}+D_{\perp}\pm 
                \sqrt{(D_{\parallel}-D_{\perp})^2-4D_{\times 1}(D_{\times 1}-D_{\times 2})}\right],
\LB{eq:growth}
\end{equation}
where $\sf k$ is the magnitude of the wave-vector of the normal-mode perturbations
proportional to $e^{\sigma t + i{\bf k}\cdot{\bf x}}$. The phase instability
becomes oscillatory when the discriminant 
\begin{equation}
-\Omega^2\equiv (D_{\parallel}-D_{\perp})^2-4D_{\times 1}(D_{\times 1}-D_{\times 2})
\end{equation}
is negative, which is possible only when the chiral symmetry is broken.

\begin{figure}
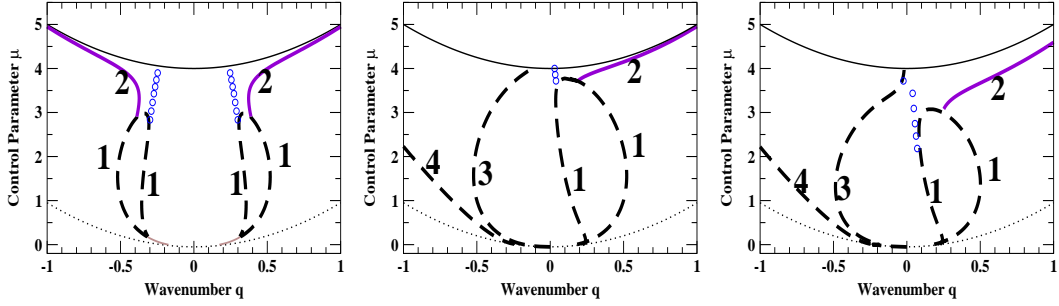

\centerline{\epsfxsize=4.5cm\epsfysize=4.0cm\epsfbox{fig2a.eps}
\hspace{0.1cm}\epsfxsize=4.5cm\epsfysize=4.0cm\epsfbox{fig2b.eps}
\hspace{0.1cm}\epsfxsize=4.5cm\epsfysize=4.0cm\epsfbox{fig2c.eps}}
\caption{
Stability diagrams for $\nu=2$, $\alpha_1=\alpha_2=\alpha_3=0$,
$\gamma=0.05$. a) $\beta=\tau=0$,  b) $\beta=-1$ and $\tau=0$, c)
$\beta=-1$ and $\tau=1$. Circles denote short-wave instabilities. Thick lines
denote long-wave instabilities: oscillatory (solid) and steady (dashed).
Dotted line denotes saddle-node bifurcation, thin solid line gives Hopf
bifurcation to oscillating hexagons.}
\LB{fig:lwsb-nomeanfl01}
\end{figure}                                             

The stability of the hexagonal pattern to general (including short-wave)
perturbations is examined by solving linearized equations similar to those
in \cite{YoRi02}. Without
rotation and the nonlinear gradient terms, the general stability boundaries
correspond to a steady bifurcation (real eigenvalues), and they always
coincide with the long-wave stability boundaries \cite{EcRi00a,YoRi02}. This
is no longer true if rotation and nonlinear gradient terms are included.
Hexagons may undergo instability via short-wave instabilities, which may be
steady or oscillatory depending on the parameters \cite{EcRi00a}. In the
following stability diagrams, we display both the stability boundary for the
long-wave perturbations (lines) and the short-wave stability boundaries
(circles).

Figs.\ref{fig:lwsb-nomeanfl01}a,b,c represent stability diagrams for
different values of $\beta$ and $\tau$ with $\gamma=0.05$ and
$\alpha_1=\alpha_2=\alpha_3=0$. Fig.\ref{fig:lwsb-nomeanfl01}a is for
$\beta=0$ and $\tau=0$,  Fig.\ref{fig:lwsb-nomeanfl01}b for $\beta=-1$
and $\tau=0$, and Fig.\ref{fig:lwsb-nomeanfl01}c for $\beta=-1$ and
$\tau=1$. 
In all the following stability diagrams 
the thick dashed curves (labeled $\bf 1$, $\bf 3$ and $\bf 4$ in
Fig.\ref{fig:lwsb-nomeanfl01}) correspond to steady phase instabilities,
whereas the thick solid lines (labeled $\bf 2$) denote the oscillatory phase
instability. The symbols denote short-wave instabilities.  As is the case
without rotation \cite{YoRi02}, mean flow renders the stability boundaries
asymmetric with respect to the band-center ($q=0$).  In addition, we find that 
the `bubble' enclosed by curve $\bf 1$ expands as the mean flow becomes
larger in amplitude  (figs.\ref{fig:lwsb-nomeanfl01}a and
\ref{fig:lwsb-nomeanfl01}b), suggesting that the mean flow tends to diminish
the importance of the oscillatory modes. For $\beta<0$, this effect appears to be more
prominent for negative $q$ than for positive $q$  and can eliminate the
oscillatory instability altogether (cf. Figs.\ref{fig:lwsb-al3-0p7}b and
\ref{fig:gsb-01}b below).

\begin{figure}[htbp]
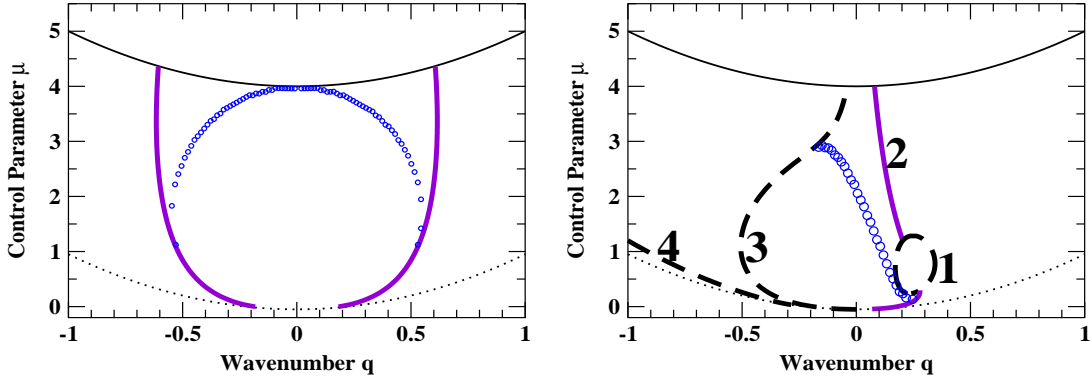

\centerline{\epsfxsize=7.0cm\epsfysize=5.0cm\epsfbox{fig3a.eps}
\hspace{0.3cm}\epsfxsize=7.0cm\epsfysize=5.0cm\epsfbox{fig3b.eps}}
\caption{
Stability boundaries for $\gamma=0$, $\alpha_{1,2}=0$, $\alpha_3=0.7$, $\nu=2$.
a) $\beta=0$, b) $\beta=-3$. Circles denote steady short-wave instability. Solid thick line
gives oscillatory long-wave instability, dashed thick line steady long-wave instability.}
\LB{fig:lwsb-al3-0p7}
\end{figure}                           

In the stability diagrams shown in Figs.\ref{fig:lwsb-al3-0p7}a,b,  we
focus on the nonlinear gradient term that is due to rotation ($\alpha_3=0.7$)
and set the other nonlinear gradient terms as well as the other rotation terms
to zero ($\gamma=0$, $\tau=0$). While the $\alpha_1$-term makes  the
stability limit of the hexagons to the mixed-mode asymmetric in $q$ and can
shift the transition from hexagons to rolls to large values of $\mu$
\cite{EcPe98}, the term corresponding to $\alpha_3$ does not affect that
amplitude instability. For $\beta=0$ (Fig.\ref{fig:lwsb-al3-0p7}a) the
long-wave instability is oscillatory, 
while the short-wave instability is steady.
When $\beta$ is decreased to $\beta=-3$ (Fig.\ref{fig:lwsb-al3-0p7}b) the
oscillatory instability is  completely suppressed for $q<0$ and replaced by
the two steady phase instabilities (labeled ${\bf 3}$ and ${\bf 4}$), whereas
for $q>0$ it is still there ($\bf 2$),  but it is mostly preempted by a steady
short-wave instability. Only for very small values of $\mu$ the oscillatory instability
remains relevant.

\begin{figure}
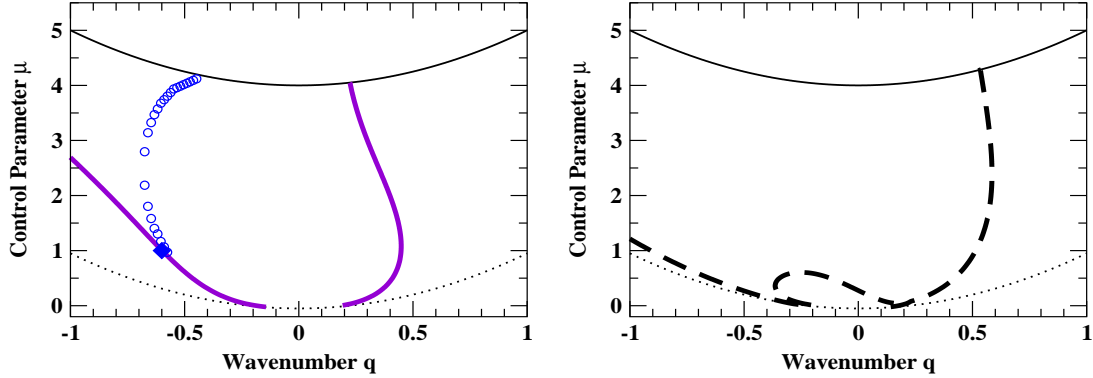

\centerline{\epsfxsize=7.0cm\epsfysize=5.0cm\epsfbox{fig4a.eps}
\hspace{0.3cm}\epsfxsize=7.0cm\epsfysize=5.0cm\epsfbox{fig4b.eps}}
\caption{
Stability boundaries for $\nu=2$, $\alpha_1=\alpha_2=0$, $\alpha_3=0.7$,
$\gamma=0.2$, $\tau=0.5$. a) $\beta=-0.1$, b) $\beta=-3$. Circles denote
steady short-wave instability. Solid thick line gives oscillatory long-wave
instability, dashed thick line steady long-wave instability. }
\LB{fig:gsb-01}
\end{figure}                        
  
If in addition to $\alpha_3$ also the cubic rotation term $\gamma$ is 
present the stability diagram becomes asymmetric even in the absence of
mean flow. This is illustrated in Fig.\ref{fig:gsb-01}a, 
which gives the stability limits for $\gamma=0.2$,
$\alpha_3=0.7$ and $\tau=0.5$, with $\alpha_1=0=\alpha_2$ and 
$\beta=-0.1$.  Note that for small $\beta$  the stability boundaries
depend very little on $\beta$ and are indistinguishable on the scale of
Fig.\ref{fig:gsb-01}a for
$\beta$ in the range $-0.2\le\beta\le 0$.   As
will be discussed in section \ref{sec:num}, however, the nonlinear evolution of the
pattern due to the instabilities can depend substantially on
$\beta$ even in this regime. The origin of the asymmetry
in $q$ can be seen from the diffusion coefficients and the growth rate
given in (\ref{eq:diff0}) and (\ref{eq:growth}), respectively. For
$\alpha_1=0=\alpha_2$ the only terms that are odd in $q$ involve the
product $\alpha_3 \omega$. For these parameter values the short-wave
instability is steady and the long-wave instability is oscillatory. The steady
long-wave instability (short segment of a dashed line) has been shifted up in
$\mu$ all the way to the  amplitude instability to oscillating hexagons  and is
preempted  by the long-wave oscillatory instability.
In Fig.\ref{fig:gsb-01}b the mean flow strength is
increased to 
$\beta=-3$.  Then the stability limits are given solely by
steady long-wave instabilities and hexagons are stable 
only over a small range in $\mu$.

\section{Defect Proliferation}
\LB{sec:num}
 
We numerically simulate equations (\ref{eq:mf1},\ref{eq:mf2}) to
investigate the non-linear dynamics ensuing from 
the linear instabilities and focus
on the combined effect of the nonlinear gradient terms and the mean flow.  
Without rotation, the mean flow couples only to the transverse phase
mode and makes it possible to discern its instability from that of the 
longitudinal phase mode \cite{YoRi02}. 
Both instabilities lead to the formation of PHD
and eventually return the wavenumber to the stable band. 
While the transient disorder generated by this process is 
quite different for the two instabilities, 
they both, in the absence of the nonlinear gradient
terms, lead to an essentially 
monotonic decay of the defect number after it has reached its initial maximum.        

With the nonlinear gradient terms included the defect dynamics can become
much more complex. In Fig.\ref{fig:num01}a,b we show the temporal evolution
of the number of dislocations for parameters corresponding to the stability
diagram shown in Fig.\ref{fig:gsb-01}a with $\mu=1$, which are
characterized by a relatively strong nonlinear gradient term that breaks the
chiral symmetry ($\alpha_3=0.7$)\footnote{While for the parameters
corresponding to fig.\ref{fig:gsb-01}b hexagons of all wavelengths are
unstable for $\mu >0.6$ numerical simulations in this regime show that the
defects arising from the instability serve as nucleation sites for rolls, which
then invade the whole system.}. Three values for the mean flow strength
are used, $\beta=0$, $\beta=-0.1$, and $\beta=-0.2$. The initial condition is
a periodic hexagon pattern with wavenumber $q=-0.6$ 
(diamond in Fig.\ref{fig:gsb-01}a) that is perturbed with a small long-wave modulation.
 
\begin{figure}
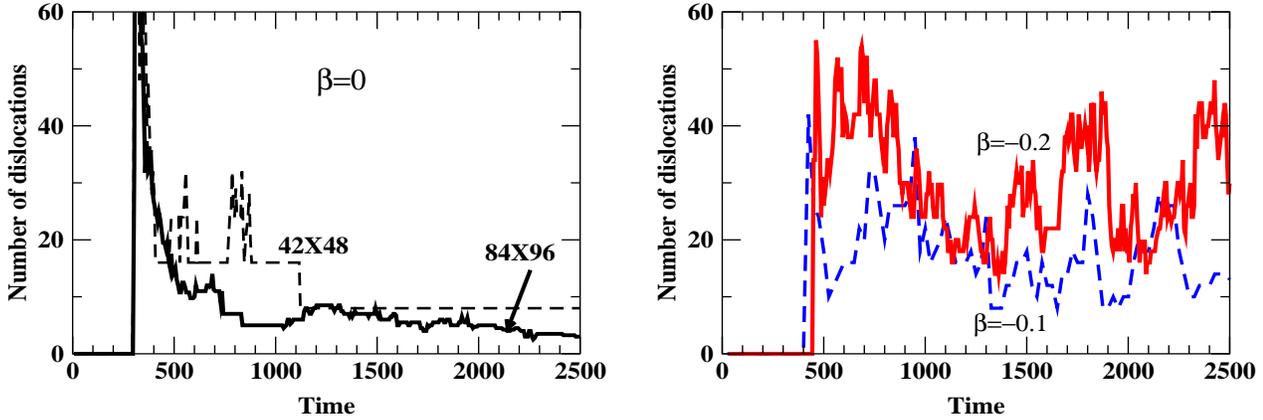

\centerline{\epsfxsize=8.0cm\epsfysize=5.5cm\epsfbox{def_count0221a.eps}
\hspace{0.5cm}\epsfxsize=8.0cm\epsfysize=5.5cm\epsfbox{def_count0221c.eps}}
\caption{Temporal evolution of the number of dislocations for different values
of $\beta$. Other parameters as in Fig.\ref{fig:gsb-01}a.
(a) $\beta=0$ for  
system size $42\times 48$ (dashed line)
and $84\times 96$ (solid line shows quarter of the number of
dislocations)
(b) $\beta=-0.1$ (dashed line) and $\beta=-0.2$ 
(solid line) for system size $42\times 48$.
}
\LB{fig:num01}
\end{figure} 
 
Since the initial wavenumber is only slightly outside  the linear stability
boundary it takes quite a long time  
until the first defects are generated.  Then
the  number of dislocations reaches a maximum very rapidly.   Without mean
flow ($\beta=0$,  Fig.\ref{fig:num01}a), it  decays subsequently to very
small values  and the final state is  presumably stationary 
(cf.  Fig.\ref{fig:num05} below). 
In contrast to the case without nonlinear gradient
terms and without rotation \cite{YoRi02}, further defects are, however,
created throughout the simulation time, in particular in the larger system of
size $84\times 96$ (solid line). 

\begin{figure}
\centerline{\epsfxsize=8.0cm\epsfysize=5.5cm\epsfbox{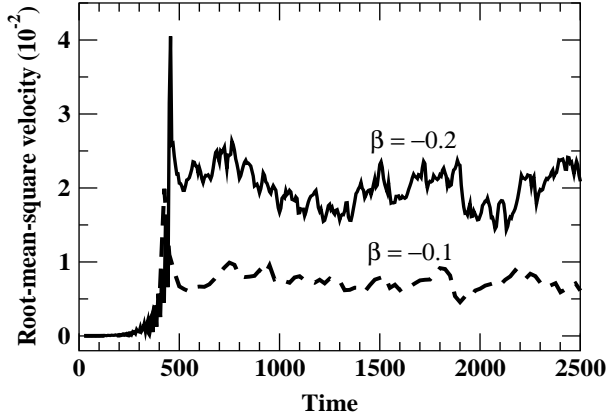}}
\caption{
Root-mean-square of the mean flow velocity 
as a function of time for $\beta=-0.1$ and $\beta=-0.2$. 
Other parameters as in Fig.\ref{fig:gsb-01}a.
}
\LB{fig:num06}
\end{figure} 
  
When turning on the mean flow in addition to the nonlinear gradient terms,
the evolution becomes more complex as shown in Fig.\ref{fig:num01}b.
Decreasing $\beta$ from  $\beta=0$ to $\beta=-0.2$ shifts the stability limits
very slightly towards less negative wavenumbers and renders the hexagons
slightly  more stable. The initial rise in the number  of dislocations is therefore
delayed to later times and the initial peak is smaller. Even though this change
in the stability limits cannot be discerned on the scale of 
Fig.\ref{fig:gsb-01}a,  the weak induced mean flow is 
sufficient to lead to strong and
persistent fluctuations in  the number of defects 
(solid line in Fig.\ref{fig:num01}b). 
In fact, somewhat smaller and slower fluctuations persist
even for $\beta=-0.1$ (dashed line).  The strong correlation between the
number  of defects and the mean-flow strength is apparent when comparing
Fig.\ref{fig:num01}b with Fig.\ref{fig:num06}, which shows the  spatial average
of the root-mean-square of the mean-flow velocity $\sqrt{<(\nabla Q)^2>}$
as a function of time.

\subsection{Induced Defect Nucleation}

A closer inspection of the defect dynamics shown in Fig.\ref{fig:num01}
reveals  that the temporal fluctuations result from the {\it induced}
nucleation of new dislocations in the vicinity of pre-existing
penta-hepta defects. One such event is shown in Fig.\ref{fig:defpro}.
The initial configuration consists of a PHD denoted by $(0,-,+)$, which
is comprised of a dislocation with negative topological charge in
mode $A_2$ and one with positive charge in mode $A_3$ (grey circles in
Fig.\ref{fig:defpro}).  Mode $A_1$ does not have any dislocations in the
vicinity of this PHD. It is, however, perturbed due to the presence of
the PHD and the zero-contour lines of the real and the imaginary part of
$A_1$ (thick solid and dashed lines) are twisted around the dislocations
in $A_2$ and $A_3$ (Fig.\ref{fig:defpro}a). This twisting reflects to
some extent the advection of this mode by the mean flow (thin lines).
Soon two dislocations appear in $A_1$  (black circles in
Fig.\ref{fig:defpro}c), which then bind with the dislocations
constituting the initial PHD to form two new PHDs, $(+.-,0)$ and
$(-,0,+)$. The new PHDs typically move apart and can induce the
nucleation of further dislocations leading to the proliferation or
multiplication of defects as sketched in Fig.\ref{fig:defproliferation}.
There the further nucleation of  a pair of dislocations in $A_2$ and
$A_3$ leads eventually to four PHDs. Such  proliferation processes have
recently been found also in simulations of Kuramoto-Sivashinsky equation
and  of Ginzburg-Landau equations without rotation and without mean flow
but with the nonlinear gradient terms corresponding to $\alpha_{1,2}$
\cite{CoNe02}.
                    
\begin{figure}
 \begin{center}
   \begin{tabular}{c c}
      (a)Time = $881.25$ & (b)Time = $887.5$ \\
      {\epsfig{file=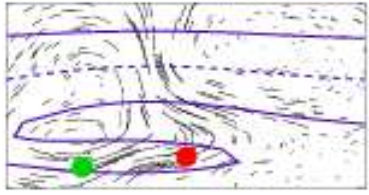,height=1.5in,width=2.8in}} &
      {\epsfig{file=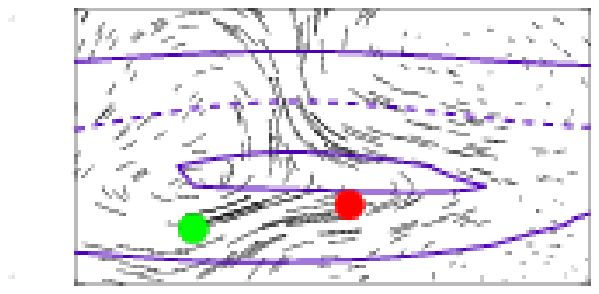,height=1.5in,width=2.8in}} \\
      (c)Time = $893.75$ & (d)Time = $900$ \\
      {\epsfig{file=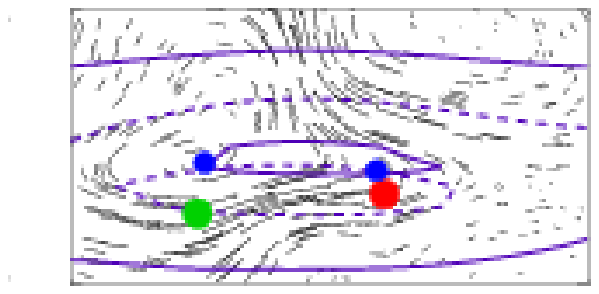,height=1.5in,width=2.8in}} &
      {\epsfig{file=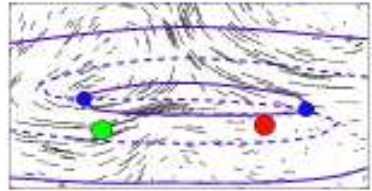,height=1.5in,width=2.8in}} \\
   \end{tabular}
 \end{center}
\caption{
Induced nucleation of dislocations by penta-hepta defects for $\beta=-0.2$
and $\mu=1$. Other parameters as in Fig.\ref{fig:gsb-01}a. Solid lines
are zero contour lines for real part of $A_1$, and dashed lines for the
imaginary part.  Grey circles are dislocations in $A_2$ and $A_3$ and the
arrows indicate the mean flow velocity field. }
\LB{fig:defpro}
\end{figure}                             
\begin{figure}
\centerline{\epsfxsize=10.0cm\epsfysize=3.0cm\epsfbox{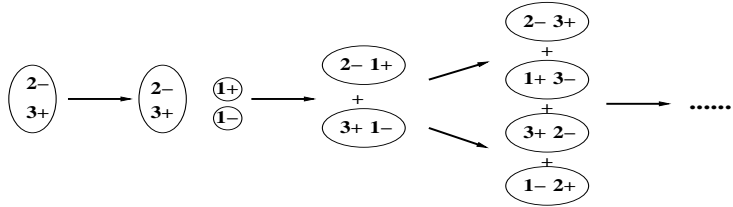}}
\caption{
Sketch of defect proliferation by induced nucleation.
}
\LB{fig:defproliferation}
\end{figure}    
   
The nonlinear gradient terms appear to be central for the induced nucleation
in that they lead to a separation of the dislocations making up the
penta-hepta defects as is illustrated in Fig.\ref{fig:defpro}a.   The
dependence of the distance between the dislocations on $\alpha_3$ is
shown in Fig.\ref{fig:dis-al3}. For $\alpha_3 \ge 0.6$ the PHD becomes
unstable through the induced nucleation of dislocations.  We find that the
dislocations are also separated if only the nonlinear  gradient terms
corresponding to  $\alpha_1$ or $\alpha_2$ are included (see also
\cite{CoNe02}). Breaking the chiral symmetry  through the $\gamma$-term
does not affect the distance between  the dislocations in a PHD pair, nor
does the mean flow.
        
\begin{figure}
\centerline{\epsfxsize=7.0cm\epsfysize=5.5cm\epsfbox{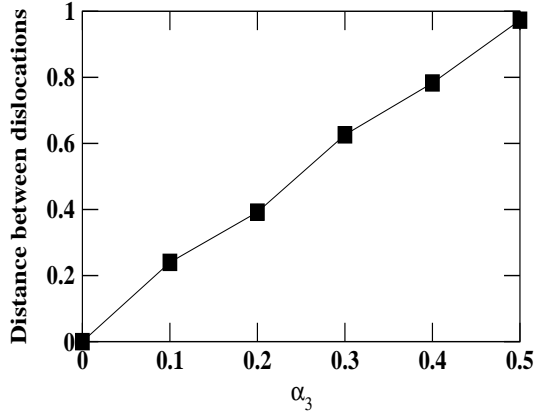}}
\caption{
Distance between dislocations within a bound PHD as a function of
$\alpha_3$ for $\beta=0$. Other parameters as in Fig.\ref{fig:gsb-01}. }
\LB{fig:dis-al3}
\end{figure}               
 
The stability of the PHD, however, depends not only on the
distance between its two constituent dislocations but also on the mean flow.
This is shown in Fig.\ref{fig:stabiPHD} where the stability limit
of PHDs is given as a function of $\alpha_3$ and $\beta$ for a
background wavenumber $q=0$. It is obtained by direct numerical
simulations of a PHD pair in a system of size $42\times 48$. Note that
for this system size there is still a small, but noticeable interaction
between the two PHDs that have to be placed in the system to
satisfy the periodic boundary conditions. In the stable regime
the two PHDs move relatively slowly towards each other and would
eventually annihilate each other, whereas in the unstable regime
the seeded PHDs nucleate new dislocations and get transformed into
different PHDs before they reach each other.

Clearly, increased mean flow greatly facilitates the induced
nucleation of dislocations rendering the PHDs much less stable
even though the distance of the constituent dislocations is
hardly affected by the mean flow. Thus, even relatively small
values of  $\alpha_3$ can be sufficient to induce nucleation if
$| \beta |={\mathcal O}(1)$. When  $\alpha_3$ is decreased below
$\alpha_3 \approx 0.1$ the induced
nucleation as described above is replaced by a different process
in which new dislocations appear not in the previously
defect-free mode but in the modes that carry already a
dislocation. This is presumably due to the  fact that with
increasing mean-flow strength the side-band stability limit of the 
periodic hexagon pattern is shifted further to the left 
(for example, see Fig.1 in \cite{YoRi02}) and comes
very close to the background wavenumber $q=0$ employed in
Fig.\ref{fig:stabiPHD}. 

\begin{figure}
\centerline{\epsfxsize=7.0cm\epsfbox{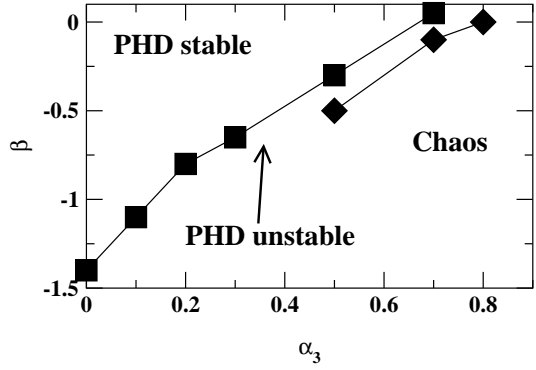}
}
\caption{
Stability limits of penta-hepta defects (squares) and limit of persistent
spatio-temporal chaos (diamonds)  as a function of mean-flow strength
$\beta$ and of $\alpha_3$. Remaining parameters are $\mu=1$, $\nu=2$,
$\gamma=0.2$, $\tau=0.5$, and $q=0$.}
\LB{fig:stabiPHD}
\end{figure}    

\subsection{Persistent PHD-Chaos}

A particularly interesting aspect of the simulations shown in
Fig.\ref{fig:num01}b is the fact that the induced nucleation of dislocations is
not merely transient as had been the case in \cite{CoNe02}, but instead
persists for the  whole duration of the simulation.  To bring out the persistence
of the dynamics more clearly, we give a detailed analysis of the pattern
evolution by measuring the local wavenumber  ${\bf q}_j({\bf x})$ of each
component,
\begin{eqnarray}
\LB{eq:def-lw}
{\bf q}_j({\bf x})&\equiv& \Re \left(\frac{-i\vec{\nabla}A_j}{A_j} \right)\;\; {\mbox{ for
$j=1$, $2$, $3$}}.
\end{eqnarray}
We extract the spatial average of the transverse and of the
longitudinal components $q_{lj}$  and $q_{tj}$ of ${\bf q}_j$, respectively, 
\begin{eqnarray}
\LB{eq:local-wn}
\overline{q_{lj}}=\overline{\hat{n}_j \cdot {\mathbf q}_j}  
\equiv \frac{\int_{\Gamma} \hat{n}_j \cdot {\mathbf q}_j \;\, d^2{\bf x}}
 {\int_{\Gamma} d^2{\bf x}},&\;\;&
\overline{q_{tj}}=\overline{\hat{\tau}_j\cdot {\mathbf q}_j} 
\equiv \frac{\int_{\Gamma} \hat{\tau}_j \cdot {\mathbf q}_j \;\, d^2{\bf x}}
{\int_{\Gamma} d^2{\bf x}}.
\end{eqnarray}
Here $\Gamma$ denotes the spatial domain. Figs.\ref{fig:num05}a,b
show the temporal evolution of the two wave-vector components
 corresponding to the evolution of
the number of defects shown in Figs.\ref{fig:num01}a,b (for
system size $42\times 48$). The rapid initial change in both
components corresponds to the period in time when the long-wave instability
first generates dislocations. For all three values of $\beta$ the
longitudinal component rapidly relaxes to a very small value
near $\overline{q_{li}}=0$  and, with the exception of small
fluctuations, stays there for the whole duration of the
simulations. Similarly, for $\beta=0$ the transverse component reaches in
a somewhat slower transient a stationary value of
$\overline{q_{tj}}\approx 1.6$. The resulting reconstructed pattern at the
final time $t=2500$ is shown in Fig.\ref{fig:num02}a. It still
contains a few PHD, which move very slowly. Compared to the
initial pattern, in which the hexagons were aligned with the 
$x$-axis, the pattern is rotated by a small amount reflecting the
change in the average wave-vector, in particular its non-zero
transverse component. 

\begin{figure}
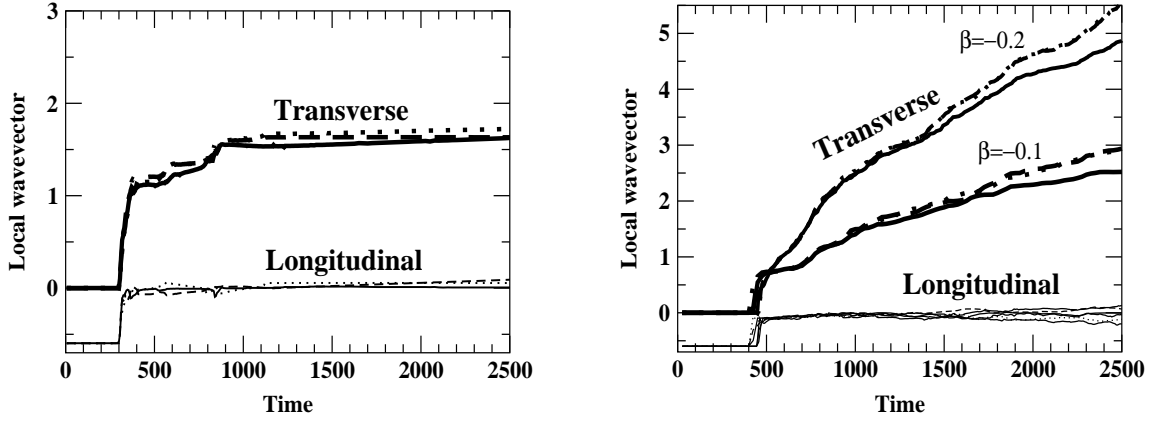

\centerline{\epsfxsize=7.0cm\epsfysize=5.5cm\epsfbox{local_wn_1004.eps}
\hspace{1cm}\epsfxsize=7.0cm\epsfysize=5.5cm\epsfbox{local_wn_041702.eps}}
\caption{
Spatially averaged local wave vectors as functions of time:  thick lines are
the transverse components $\overline{q_{tj}}$ and thin lines are the
longitudinal components $\overline{q_{lj}}$ of the wave-numbers. Solid lines
are for $A_1$, dashed lines for $A_2$ and dotted lines for $A_3$.
Panel (a) is for $\beta=0$ and panel (b) is for $\beta=-0.1$ and
$\beta=-0.2$.}
\LB{fig:num05}
\end{figure}  

For $\beta=-0.1$ and $\beta=-0.2$ the number of PHDs keeps
fluctuating throughout the simulation. The corresponding
evolution of the average wave-vector is shown in
Fig.\ref{fig:num05}b. As is the case for $\beta=0$, during the
initial phase of defect generation the longitudinal component
relaxes rapidly to $\overline{q_{lj}}=0$ while the transverse
component reaches a value of $\overline{q_{tj}}\approx 1$ during
that phase. Thereafter, the longitudinal component remains near
$0$ whereas the transverse component keeps growing at a rate
which increases with decreasing (negative) $\beta$. In
the reconstructed pattern, shown in
Fig.\ref{fig:num02}b at $t=2500$, this manifests itself in a reduced overall
wavelength of the hexagons and, more importantly, an increased
rotation of the pattern.  Following the reconstructed pattern
over time one can see that on average it is steadily rotating in
a counter-clockwise fashion. 

\begin{figure}
\centerline{\epsfxsize=7.0cm\epsfysize=7.0cm\epsfbox{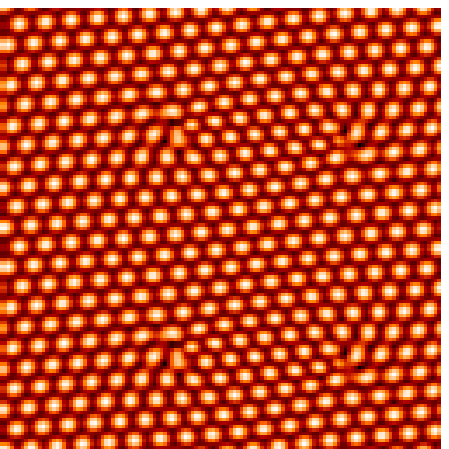}
\hspace{1cm}\epsfxsize=7.0cm\epsfysize=7.0cm\epsfbox{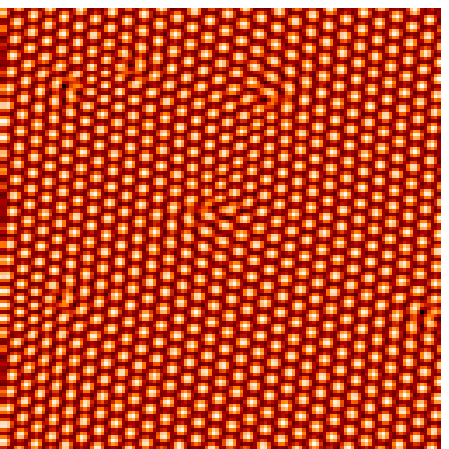}}
\caption{
Reconstructed hexagonal patterns at the end of the 
simulations ($t=2500$) for $\beta=0$ on the left and $\beta=-0.1$ on the right.
}
\LB{fig:num02}
\end{figure}      

To interpret the evolution shown in Fig.\ref{fig:num05}b it is useful
to write $A_i({\bf x},t)$ as
\bea
A_j={\mathcal A}_j({\bf x},t) e^{i q_t \hat{\tau}_j\cdot {\bf x}}, 
\qquad j=1,2,3.\LB{e:Atrans}
\eea 
For constant ${\mathcal A}_j$ this corresponds to a hexagon with
non-vanishing transverse wave-vector components that are equal in all three
modes. Insertion of (\ref{e:Atrans}) into (\ref{eq:mf1},\ref{eq:mf2}) shows that
the solution and its stability does not depend on $q_t$. In particular, the
contributions from $\alpha_2$ cancel each other. This independence of
$q_t$ reflects the isotropy of the underlying physical system and the
leading-order approximation of the critical circle by its tangents in the
direction of $\hat{\tau}_j$ at each of the three critical wave-vectors
corresponding to the modes $A_j$. This suggests that the  steady increase of
$\overline{q_{tj}}$ with fixed $\overline{q_{lj}}$ should be interpreted as the
representation of a rotation of the  physical pattern (at fixed magnitude of the
wave-vector) within the approximation of the Ginzburg-Landau equations
(\ref{eq:mf1},\ref{eq:mf2}). A similar phenomenon had been observed for
hexagons with broken chiral symmetry in the absence of mean flow
\cite{EcRi00}. In that system it had been found that if the operator
$\hat{n}_j\cdot \nabla$ was replaced by the Newell-Whitehead-Segel
operator \cite{NeWh69,Se69}  or by the Gunaratne operator \cite{GuOu94}
the steady growth of the wave-vector components did not follow $\hat{\tau}_j$
but instead followed the respective lines in Fourier space along which the
growth rate of perturbations of the basic state is constant within these two
different approximations \cite{EcRiunpub}.  

Given the above discussion, it is appropriate to consider within the
approximation (\ref{eq:mf1},\ref{eq:mf2}) all hexagon solutions with
$q_{lj}=0$ as having a wave-vector at the band-center, with $q_{tj}$
indicating the orientation of the pattern in space. This raises the
question why new dislocations are generated persistently even though the
background  wavenumber is in the band center where the  periodic 
hexagon patterns are linearly stable? In \cite{CoNe02} it had been found
that without mean flow the PHDs can be unstable even if the wavenumber
of the background hexagon pattern is inside the stable band, but away
from the  band center.  As was shown in  Fig.\ref{fig:stabiPHD} above,
in the presence of mean flow PHDs can be unstable even at the band
center. We have not investigated whether there are parameter regimes for
which the PHDs are unstable at all background wavenumbers. While
instability at the band center is consistent with the persistence of
spatio-temporal chaos since the background wavenumber of the chaotic
state is $q=0$, it is not sufficient. This is indicated in
Fig.\ref{fig:stabiPHD}, where also the persistence limit for the chaotic
state is shown. Below the squares the PHD's are unstable but chaos only
persists for values of $\beta$ below the diamonds. In the parameter
regime between these two lines defects are being created intermittently
during the  transient towards the stationary state, which leads to
fluctuations in the defect number,  but the defect creation rate is too
small compared to the annihilation rate to sustain the chaos.

To illustrate explicitly how the instability of a PHD at the band center can
lead to persistent PHD chaos, we performed simulations that start from
hexagons at the band center ($q=0$) with a pair of PHDs as a seed for the
nucleation process. The resulting evolution of the number of defects and of the
spatially averaged transverse wave-vector component is shown in
Figs.\ref{fig:alpha05}a,b,  respectively.  Compared to the simulations shown
in  Fig.\ref{fig:num02}b, the value of $\alpha_3$ is reduced to
$\alpha_3=0.5$. As a consequence even for $\beta=-0.2$ no new
dislocations are generated. For $\beta=-0.4$ there is an initial volley of
induced nucleation, but eventually the defects annihilate each other again
completely. Only for $\beta \le -0.5$ indications 
for persistent nucleation are seen. 
This illustrates that the instability of PHD is not a sufficient condition for
persistent chaos; it is necessary that
creation balances annihilation for some finite average number of PHDs.

Worth noting is the precipitous decline in the defect number for $\beta=-0.6$
near $t=900$. At that time the transverse wave-vector component reaches 
values close to the maximal spatial resolution for the number of modes used
($128\times 128$). More detailed studies of the case with $\alpha=0.7$ and
$\beta=-0.2$ have shown that if the number of modes is increased the
suppression of the induced nucleation is delayed until larger values of 
$\overline{q_{tj}}$ are reached. Thus, the saturation of  $\overline{q_{tj}}$ for
$\beta=-0.5$ in Fig.\ref{fig:alpha05}b and the associated end of the chaotic
activity is a numerical artifact. Note, however, that the chaotic defect states
reported here have lasted substantially longer than the transient states found
in \cite{CoNe02}.

\begin{figure}
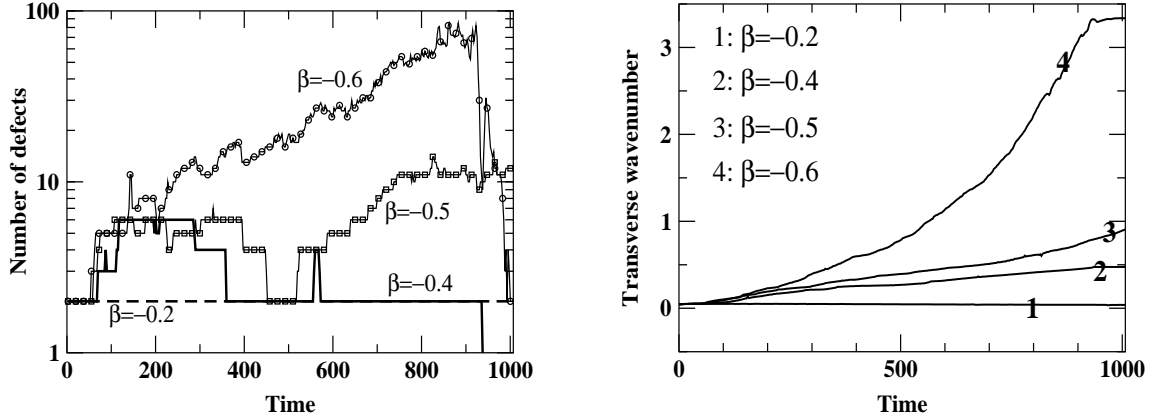

\centerline{
\epsfxsize=7.0cm\epsfysize=5.5cm\epsfbox{rdefpos_beta_al3_0p5.eps}
\hspace{1cm}\epsfxsize=7.0cm\epsfysize=5.5cm\epsfbox{lcwn_beta_al3_0p5a.eps}
}
\caption{
a) Temporal evolution of number of dislocations for $\alpha=0.5$ and various
values of $\beta$. 
Other parameters as in Fig.\ref{fig:num01}. b) Evolution of the
spatially averaged transverse wave vector component $\overline{q_{tj}}$.   }
\LB{fig:alpha05}
\end{figure} 
 
Since the defect nucleation persists even for wavenumbers near the
band-center it is not surprising that similar chaotic states are reached when
the stability limit for periodic hexagons is crossed on the large-$q$ side. It
should be noted that in simulations that started from a slightly perturbed
hexagon pattern beyond the low-$q$ stability limit no persistent nucleation
was found when the wavenumber was too far in the unstable regime. In that
case only a very large number of defects was generated in the initial phase,
which then quickly annihilated each other.
 
Of course, the steady increase of the transverse wave-vector component
apparent in Fig.\ref{fig:num05}b implies that after a finite amount of time the
magnitude of the transverse component $q_{tj}$ becomes of the same order
as the critical wavenumber. At that point the Ginzburg-Landau equations
clearly will have become invalid, since they are based on the approximation
that the amplitudes $A_j$ vary only slowly compared to the critical
wavelength. We have therefore also simulated a modified Swift-Hohenberg
equation that  corresponds to the same amplitude equations with  coefficients
corresponding to the values used in the simulations presented here. As
expected, we find that the hexagon pattern keeps rotating on average as the 
defect proliferation and annihilation continue indefinitely \cite{YoRi02b}.

In general, the other nonlinear gradient terms corresponding to $\alpha_1$
and  $\alpha_2$ have to be taken into account as well. We have performed
simulations for selected parameter values of $\alpha_1$ and $\alpha_2$ that
correspond, for instance, to realistic values for surface-tension-driven
B\'{e}nard-Marangoni convection ($\alpha_1=0.3$ and $\alpha_2=1.1$,
corresponding to a Prandtl number $\sim 10$ in a single-fluid model
\cite{BrVe98}; $\alpha_1=-0.1$ and $\alpha_2=0.4$ corresponding
to a two-fluid model a water-air layer \cite{GoNe97}).  While
we find induced nucleation even in the absence of rotation and without
$\alpha_3$,  persistent spatio-temporal chaos arises  only if $\alpha_3 \ge
0.8$ in both cases ($\beta=-0.2$).

  \YY{$\beta=-0.2$}

\section{Effects of non-linear gradient terms on PHD motion}
\LB{sec:def}

The motion of individual PHD's within the lowest-order Ginzburg-Landau
equations has been studied in detail in \cite{Ts95,Ts96}, where in extension
of the results for dislocations in roll patterns
\cite{BoPe88} the dependence of the velocity of
a PHD on the background wave-vectors of the three hexagon modes has
been determined semi-analytically. The effect of the mean flow on the defect
motion has  been discussed in some detail in \cite{YoRi02}. Here we turn to
the impact of the other non-variational terms, specifically the non-linear gradient
terms, on the motion of the PHD's. 

In our analysis we closely follow the approach of \cite{Ts96}. We factor out 
the background wave-vectors of the three hexagon modes by writing
$A_j\equiv a_j e^{i{\bf q}_j\cdot {\bf x}}$.    Assuming a constant defect
velocity ${\mathbf V}$, the time derivative in (\ref{eq:mf1}) is replaced by
$-{\mathbf V}\cdot\nabla$.  We then project (\ref{eq:mf1}) onto   the two
translation modes, i.e. for each $j$ we multiply (\ref{eq:mf1}) by $\partial_x
a_j^*$ and $\partial_y a_j^*$, respectively, and add all three equations and
their complex conjugates.  The resulting projection can be written as
\begin{eqnarray}
\LB{e:mob1}
I_{xx}V_x + I_{xy}V_y = F, &\;\;\;\;\;\;\;\;&
I_{yx}V_x + I_{yy}V_y = G,
\end{eqnarray}
where the components of the mobility tensor are given by 
\begin{eqnarray}
\LB{e:mob2}
I_{xx} \equiv &&\left<\sum_{i=1}^3|\partial_x a_i|^2\right> ,\;\;
I_{yy} \equiv \left<\sum_{i=1}^3|\partial_y a_i|^2\right>,\nonumber \\
I_{xy}=&&I_{yx}= \frac{1}{2}\left<\sum_{i=1}^3\partial_xa_i\partial_ya_i^*+c.c.\right>,
\end{eqnarray}
and $<\cdots>\equiv\int\int \cdots dxdy$ denotes the integral over the domain.
The terms on 
the right-hand-side of equation (\ref{eq:mf1}) contribute to  $F$ and $G$ with
\begin{eqnarray}
\LB{e:mob3}
F&=&-2\gamma\left<\sum_{i=1}^3|a_i|^2\partial_x|a_{i+1}|^2\right> 
    -i\left<\sum_{i=1}^3q_i\partial_xa_i^*(\niig)a_i-c.c.\right> \nonumber \\
 && -i\frac{\beta}{2}\left<\sum_{i=1}^3\partial_xa_i^*a_i(\tiig)Q-c.c.\right> \nonumber \\
 && -i\frac{\alpha_1}{2}\left<\sum_{i=1}^3\partial_xa_i^*
    \left[a_{i-1}^*(\nkkg)a_{i+1}^*+ a_{i+1}^*(\njjg)a_{i-1}^*\right]-c.c.\right> \nonumber\\
 && -i\frac{\alpha_2}{2}\left<\sum_{i=1}^3\partial_xa_i^*
    \left[a_{i-1}^*(\tkkg)a_{i+1}^*- a_{i+1}^*(\tjjg)a_{i-1}^*\right]-c.c.\right> \nonumber\\
 && -i\frac{\alpha_3}{2}\left<\sum_{i=1}^3\partial_xa_i^*
    \left[a_{i-1}^*(\nkkg)a_{i+1}^*- a_{i+1}^*(\njjg)a_{i-1}^*\right]-c.c.\right> \nonumber\\
 &\equiv& \gamma F_{\gamma}+q_0F_q+\beta F_{\beta}+\alpha_1F_1+\alpha_2F_2+\alpha_3F_3.  
\end{eqnarray}
$G$ is defined analogously to $F$ with $\partial_x a_i^*$  in front of the
square brackets in each of the integrals replaced by $\partial_y a_i^*$.  The
projection $q_i$ of the wave-vector ${\bf q}$ onto ${\hat n}_i$  is given by
$q_i={\bf q}_i\cdot{\hat n}_i$. To leading order the transverse components
${\bf q}_i \cdot \hat{\tau}_i$ of the wave-vectors do not affect the velocity of
the defect.  Here we only focus on situations where $q_i\equiv q_0$ is the same
for all three amplitudes.   Note, that the other terms in (\ref{eq:mf1}) do not
contribute to (\ref{e:mob3}), which can be seen by integrating by parts and using
the appropriate boundary conditions.  

In the absence of the non-potential terms 
($\gamma=\beta=\alpha_1=\alpha_2=\alpha_3=0$) the PHD is stationary for
$q_0=0$ and the two dislocations making up the PHD are located at the
same position. Without loss of generality, we assume that the PHD consists of
a pair of dislocations in $A_2$ and $A_3$. Treating the non-potential terms
perturbatively it is sufficient to use the potential solution for the stationary
defect to evaluate the inhomogeneous terms $F$ and $G$. Far away from the
defect core the pattern is well described by the phase equations; rewriting
the demodulated amplitude as $a_j=\rho_j(r,\phi)e^{i\theta_j(r,\phi)}$, where
$r$ is the radial distance from the defect core and $\phi$ is the azimuthal
angle around the core, one obtains then \cite{PiNe93} $\rho_i(r
\rightarrow\infty,\phi)\rightarrow\rho_0$ and
\begin{eqnarray}
\theta_1 &=&\frac{1}{2\sqrt{3}}(1-\cos(2\phi)),\\
\theta_2 &=& \phi-\frac{1}{2\sqrt{3}}\left[\frac{1}{2}+\cos(2\phi-\frac{2\pi}{3})\right],\\
\theta_3 &=&-\phi-\frac{1}{2\sqrt{3}}\left[\frac{1}{2}+\cos(2\phi+\frac{2\pi}{3})\right].
\end{eqnarray}
This solution is utilized for the far-field contribution to the integrals for $F$
and $G$. The numerical solution of the Ginzburg-Landau equation 
(\ref{eq:mf1},\ref{eq:mf2}) is used for the evaluation of the integral in the core
region.  It can be shown that the far-field contribution to the integrals $G_1$, 
$G_2$, and $F_3$ is zero,  while that to the integrals $F_1$, $F_2$, and
$G_3$ is non-zero. For example,  for $G_1$ the angular integral involving
$\theta_1$ vanishes, and those involving $\theta_2$ and $\theta_3$ cancel
each other, leading to a vanishing contribution to $G_1$ from the far field. Numerically, we
also found that the core integrals for $G_1$, $G_2$, and $F_3$ vanish; they
are at least $100$ times smaller than the core integrals for $F_1$,  $F_2$,
and $G_3$.  In the following we ignore $F_{\gamma}$ and $G_{\gamma}$;
similar to the integrals $G_1$, $G_2$, and $F_3$, the integrals
$F_{\gamma}$ and $G_{\gamma}$ vanish in the far-field and amount to very
small values when evaluated numerically around the core. Thus, we
conclude that to leading order in the non-potential terms (and the
wavenumber $q_0$) the velocity of an individual PHD is well approximated
by 
\begin{eqnarray}
\LB{e:mob5}
\left(\begin{array}{cc}I_{xx}&I_{xy}\\
                       I_{yx}&I_{yy}
      \end{array} \right)\left(\begin{array}{c} V_x\\V_y\end{array}\right) &=&
q    \left(\begin{array}{c}F_q\\G_q\end{array}\right)+
\beta\left(\begin{array}{c}F_{\beta}\\G_{\beta}\end{array}\right)+
\left(\begin{array}{c}\alpha_1F_1+\alpha_2F_2\\\alpha_3G_3\end{array}\right).
\end{eqnarray}

To evaluate the mobility tensor on the left-hand side of (\ref{e:mob5}) it is not
sufficient to insert the solution for the stationary defect in (\ref{e:mob2}) since
it leads to  integrals that diverge in the far field. To regularize this singularity
the solution for the moving defect has to be used, at least in the far field
\cite{SiZi81a,BoPe88,Ts96}. To leading order, the non-potential terms can
be neglected in the mobility tensor. Then its off-diagonal terms are much
smaller than the diagonal terms \cite{Ts96}. Using the numerically determined
defect solution of the full equations (\ref{eq:mf1},\ref{eq:mf2}), we also find
$I_{xy}$ to be much smaller than either $I_{xx}$ or $I_{yy}$ (by a factor of
$100$) when we only compute the integral within a box enclosing the defect
core and neglect the far-field contribution.

The effect of the non-linear gradient terms can now be summarized as
follows: since the off-diagonal terms in the mobility tensor are small,  the
contributions of $\alpha_1$ and $\alpha_2$  to the velocity of a PHD with
dislocations in amplitudes $A_2$ and $A_3$ are in the
$\hat{n}_1$-direction (``glide"), while $\alpha_3$ causes such a PHD to
travel in the ${\hat \tau}_1$-direction (``climb"). Furthermore, like the  mean flow
the nonlinear  gradient terms contribute to a shift of the wavenumber at which
the defect is stationary away from the band center $q=0$.
 
The above conclusion is confirmed in direct numerical simulations of
(\ref{eq:mf1},\ref{eq:mf2}).  In Fig.\ref{fig:def-vel} we plot the velocity of
a PHD with charges $(0,-,+)$ for $\mu=1$, $q_0=0$, $\gamma=0.2$, and $\beta=0$ as a function
of the strength of the nonlinear gradient terms. As expected from the
discussion  of (\ref{e:mob5}), if only $\alpha_1$ or $\alpha_2$ are non-zero
the defect glides, whereas it climbs if only $\alpha_3$ is present. The velocities
change sign if the charges of the PHD are reversed. 

\begin{figure}
\centerline{
\epsfxsize=7.0cm\epsfysize=5.5cm\epsfbox{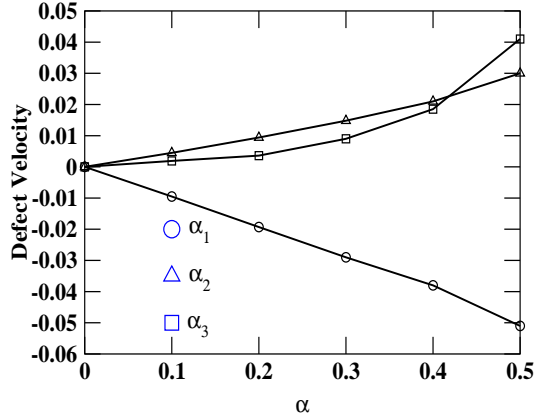}}
\caption{
Defect velocity due to $\alpha_{1,2}$ (glide) and due to $\alpha_3$ (climb),
respectively, for $q_0=0$, $\mu=1$, $\gamma=0.2$, $\beta=0$, and $\nu=2$. }
\LB{fig:def-vel}
\end{figure}       

\begin{figure}
\centerline{\epsfxsize=7.0cm\epsfysize=5.5cm\epsfbox{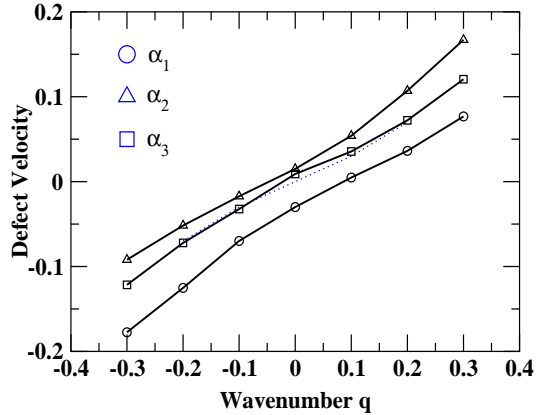}}
\caption{
Dependence of the defect velocity on the wavenumber $q$
for $\mu=1$, $\nu=2$ and $\beta=0$. $\alpha_i=0$ except as indicated: $\alpha_1=0.3$
(circles), $\alpha_2=0.3$ (triangles), $\alpha_3=0.3$ (squares).
Dotted line is for the variational case $\alpha_1=\alpha_2=\alpha_3=0$.
}
\LB{fig:def-velq}
\end{figure}  
 
Fig.\ref{fig:def-vel} indicates that the linear dependence of the
velocity   on the $\alpha_i$ is restricted to a range $|\alpha_i| \le 0.2$. It is
found,  however, that the direction of the defect motion remains the same for
$\alpha_i$ outside that range, i.e. $\alpha_1$ and $\alpha_2$ cause defects
to glide while defects climb due to $\alpha_3$.  In comparison, in previous
numerical simulations we found that the mean flow causes the PHD's to
perform a combined climb-glide motion \cite{YoRi02}.  This indicates that
both $F_{\beta}$ and $G_{\beta}$ are non-zero in equation (\ref{e:mob5}).
Fig.\ref{fig:def-velq} shows the dependence of the defect velocity on the
wavenumber $q$. For $\alpha_3 \ne 0$, the defect
performs a mixed climb-glide motion if the wave number is not at the band center, i.e.
if $q \ne 0$.

\section{Conclusion}
\LB{sec:con}

The variational character of the minimal Ginzburg-Landau equations for
steady hexagon patterns can be broken in various ways. Quite generally, at
cubic order two nonlinear gradient terms arise \cite{Br89,KuNe95}. If the
chiral symmetry of the system is broken (e.g. by rotation) a third nonlinear
gradient term is possible \cite{EcRi00}. In Rayleigh-B\'enard and in
Marangoni convection the mean flow driven by deformations of the pattern
introduce a further non-variational term in the form of a non-local coupling
of the roll modes
(e.g. \cite{SiZi81,GoNe95a}). In this paper we have studied the combined effect of
the mean flow and of rotation (including the respective nonlinear gradient
term) on the stability and dynamics of hexagon patterns as well as on the
stability and motion of their penta-hepta defects. 
 
Rotation induces a coupling of the two long-wave phase modes that can
generate a long-wave oscillatory instability, which in the absence of
nonlinear gradient terms is, however, usually screened by a 
steady short-wave instability \cite{EcRi00}. The mean flow is only driven by
the transverse phase mode \cite{YoRi02}. Our results indicate that it
suppresses the rotation-induced coupling of the phase modes that leads to
oscillatory  behavior. 

The various non-variational terms influence the motion of penta-hepta 
defects in different ways. While the nonlinear gradient terms that preserve
the chiral symmetry contribute to a gliding motion of the defects, 
the nonlinear gradient term introduced by rotation induces a climbing motion.
This is to be contrasted with the effect of the mean flow, which leads to 
a mixed climbing-gliding motion.
 
The most interesting result of this paper is associated with the fact that the
nonlinear gradient terms lead to a separation of the two dislocations making
up a penta-hepta defect and can destabilize it \cite{CoNe02}.  More
specifically, in simulations we find that penta-hepta defects induce the
nucleation of new dislocations in the defect-free amplitude. Even a weak
mean flow can enhance this instability significantly. Moreover, with a
sufficiently strong  nonlinear gradient term that breaks the chiral symmetry
($\alpha_3$) the induced nucleation can lead to an ever-increasing
transverse wavevector component of the hexagon patterns. As in the case
without mean flow \cite{EcRi00},  we interpret this growth, which eventually
leads to the break-down of the Ginzburg-Landau equations, as the signature
of a persistent  precession of the disordered pattern on average. In Fourier
space, the wave-vector spectrum of such a precessing pattern would drift
along the critical circle. In the lowest-order Ginzburg-Landau equations
used here the critical circle is replaced by its tangents at each of the three
modes making up the hexagons, which are transverse to the respective
wave-vectors. We expect therefore that in this regime the physical system
would exhibit persistent penta-hepta defect chaos driven by induced
nucleation.

To overcome the limitations of the Ginzburg-Landau equations, we are
currently investigating penta-hepta defect chaos using a suitably modified
Swift-Hohenberg-type equations coupled to a mean flow \cite{YoRi02b}. As
expected from the simulations of the Ginzburg-Landau equations  presented
in the present paper, the penta-hepta defect chaos persists and on average the
disordered hexagon patterns precess indefinitely. As in the Ginzburg-Landau
case, the chaotic state is due to the induced nucleation of penta-hepta
defects, which is apparently only possible when the nonlinear gradient terms
are included. These simulations also indicate that the induced nucleation is
not contingent on the oscillatory sideband instability; rather it is the
separation of the dislocations in a penta-hepta defect that renders them
unstable and induces the nucleation. To obtain persistent chaos the resulting
defect proliferation has to be strong enough compared to the annihilation 
rate and apparently the chiral symmetry has to be broken.  

We wish to acknowledge useful discussion with A. Golovin, A.
Nepomnyashchy, and L. Tsimring.  This work is supported by the
Engineering Research Program of the Office of Basic Energy Sciences at
the Department of Energy (DE-FG02-92ER14303), by a grant from NASA
(NAG3-2113), and  by a grant from NSF (DMS-9804673).  YY acknowledges
computation support from the Argonne National Labs and the DOE-funded
ASCI/FLASH Center at the University of Chicago.


\end{document}